\documentclass[onecolumn,11pt]{article}
\usepackage{graphics}
\usepackage[pdftex]{graphicx}
\usepackage{amsmath,amsthm, amssymb}
\usepackage{hyperref}
\newtheorem{defn}{Definition}
\newtheorem{theorem}{Theorem}

\newtheorem{result}{Result}
\newtheorem{rem}{Remark}
\begin{document}
	\title{\textbf{\Large A new integrated likelihood for estimating population size in dependent dual-record system\footnote{This is the pre-peer reviewed version of the following article: \textit{Chatterjee, K. and Mukherjee, D. (2018). A New Integrated Likelihood for Estimating Population Size in Dependent Dual-record System. Canadian Journal of Statistics. 46, 577-592}, which has been published in final form at \href{http://dx.doi.org/10.1002/cjs.11477}{[Weblink]}. This article may be used for non-commercial purposes in accordance with Wiley Terms and Conditions for Use of Self-Archived Versions.}}}
	\author{Kiranmoy Chatterjee\thanks{Department of Statistics, Bidhannagar College, Kolkata, India; E-mail: \emph{kiranmoy07@gmail.com}} \\
		Diganta Mukherjee
		\thanks{Sampling and Official Statistics Unit, Indian Statistical Institute, Kolkata-700108, India;}}
	\date{}
	\maketitle

\begin{abstract}
Efficient estimation of population size from dependent dual-record system (DRS) remains a statistical challenge in capture-recapture type experiment. Owing to the nonidentifiability of the suitable Time-Behavioral Response Variation model (denoted as $M_{tb}$) under DRS, few methods are developed in Bayesian paradigm based on informative priors. Our contribution in this article is in developing integrated likelihood function from model $M_{tb}$ based on a novel approach developed by Severini (2007, Biometrika). Suitable weight function on nuisance parameter is derived under the assumption of availability of knowledge on the direction of behavioral dependency. Such pseudo-likelihood function is constructed so that the resulting estimator possess some desirable properties including invariance and negligible prior (or weight) sensitiveness. Extensive simulations explore the better performance of our proposed method in most of the situations than the existing Bayesian methods. Moreover, being a non-Bayesian estimator, it simply avoids heavy computational effort and time. Finally, illustration based on two real life data sets on epidemiology and economic census are presented.
\end{abstract}


\section{INTRODUCTION}
Dual-record System (DRS) is a special type of data-structure obtained from a capture-recapture type experiment, especially designed for estimating a specified population size, say $N$, from two sampling occasions. Federal agencies are generally interested to know the actual size of a specified population. Often census operations fail to extract the true knowledge on the size of the population for various reasons. Therefore, either any other contemporary census counting data is used or another survey is conducted independently after the census operation in order to estimate the true size $N$. Application of this technique is common in various interdisciplinary platforms, such as epidemiological study, socio-economic census, study of episodic events, etc. To estimate $N$ from these two lists of information (each of them is supposed to be incomplete), idea from a popular technique, called capture-recapture, is borrowed. Wolter (1986) sketched different capture-recapture models for \emph{N} from DRS based on the pioneering work of Otis et al. (1978) who introduced several likelihood models for different plausible situations. In practice for homogeneous group, model $M_t$ has received much attention from both the frequentist and Bayesian statisticians due to its simplicity. $M_t$ accounts for time(\textit{t}) variation effect and assumes causal independence between the sources of information. In the capture-recapture literature on model $M_t$, several Likelihood (e.g. Bishop et al., 1975; Huggins, 1989), pseudo-likelihood (e.g. Bolfarine et al., 1992; Chatterjee and Mukherjee, 2016a) and Bayesian methods (e.g. George and Robert, 1992; Xu et al., 2014) are available. In this context, the popular Lincoln-Petersen estimator is identical with the estimator derived from conditional likelihood (Chatterjee and Mukherjee, 2016a). But the specific assumption of causal independence in $M_t$ may seriously mislead in most of the situations for human population, especially when capture probabilities also vary with behavioral response (Chandrasekar and Deming, 1949). When both the time (\textit{t}) variation effect and behavior response (\textit{b}) effect acts together, model $M_{tb}$ is appropriate. Moreover, this model can be treated as most general statistical form of capture-recapture model for homogeneous population. Gosky and Ghosh (2011) found the model $M_{tb}$ as the most robust model in estimating $N$ based on comparative simulation study in Bayesian paradigm over all the models proposed in Otis et al. (1978). The underlying behavior response effect (say, $\phi$) classifies a given population as \textit{recapture prone} (when $\phi>1$) or \textit{recapture averse} (when $\phi<1$). Usually, demographic study exhibits \textit{recapture prone} type dependence. On the contrary, study on drug abused population usually reveals that underlying list-dependence is negative, i.e., drug abused population is \textit{recapture averse}. However, Otis et al. (1978) addressed the non-identifiability problem related to this model. Lloyd$'$s (1994) martingale approach, Chao et al.$'$s (2000) quasi-likelihood approach and Yang and Chao$'$s (2005) univariate Markovian method successfully solve the nonidentifiability for number of sampling occasions ($T$) strictly more than two. Lee et. al. (2003) successfully developed a full Bayesian technique with little informative prior but their demonstration is in the spirit of large number of sampling occasions, which is seldom used for human population. Later, Wang et al. (2015) proposed a hierarchical Bayesian $M_{tb}$ model for multiple lists with the assumption that the odds of recapture bears a constant relationship to the odds of initial capture. Capture-recapture type experiment with $T=2$ is commonly exercized in most of the applications for human population and estimation of $N$ is found to be an challenging task in the presence of causal dependence between sampling occasions. Chatterjee and Mukherjee (2016b) discusses some issues related to full Bayes method with non-informative prior in this context. They also developed some empirical Bayes strategies in DRS considering the present problem in a missing data framework. Generally, in Bayesian paradigm, difficulty may arise as the resulting estimator for $N$ may be very sensitive to the choice of prior(s). However, Nour (1982) proposed an estimator in DRS with the assumption equivalent to \textit{recapture proneness} but  avoiding Bayesian technique. Literature (Nour, 1982; Chatterjee and Mukherjee, 2016b) reveals that if correct available directional knowledge on $\phi$ is applied, inference on $N$ could be improved certainly in both of Bayesian and non-Bayesian paradigms for model $M_{tb}$. However, an efficient classification strategy (recapture proneness or aversion) of the given population in terms of the behavioral nature is proposed in Chatterjee and Mukherjee (2016c). Appropriateness and the challenging identifiability problem of the present model under DRS motivate us to consider the problem of $N$ estimation in the present paper. Here, we proposed a novel integrated likelihood method as a suitable non-Bayesian strategy to meet our goal particularly when the underlying population is correctly known as recapture prone or averse.

All the model parameters in $M_{tb}$ except the interest parameter $N$ are regarded as nuisance parameters (say, $\psi\in\Psi$). In these contexts, some useful likelihood-based inference procedures through the construction of pseudo-likelihood functions by eliminating the nuisance parameters are discussed in Severini (2000). This elimination of nuisance parameter may be done by maximization (\textit{profile likelihood}) or conditioning (\textit{conditional likelihood}) or integrating the likelihood function over $\Psi$ with respect to some weight function (\textit{integrated likelihood}). Integrated likelihood has an advantage that it always exists unlike other pseudo-likelihood methods.
Salasar et al. (2014) analysed integrated likelihood approach with uniform and Jeffrey's prior for eliminating nuisance parameters in $M_{t}$. In integrated likelihood method the main challenge is to choose a suitable prior weight function on the nuisance parameters. Severini (2007) presents a novel approach for selecting a weight function so that the resulting integrated likelihood is useful for non-Bayesian inference and also posses some nice statistical properties. Recently, Chatterjee and Mukherjee (2016a) has developed an integrated likelihood for $M_t$, with the help from Severini (2007). In this article, we extend the work of Chatterjee and Mukherjee (2016a) for our present complex model $M_{tb}$-DRS. But the main challenge here is to successfully overcome the current model identifiability problem by suitably choosing informative priors so that some desirable properties hold. In summary, this article is framed to provide an alternative or supplement to the few existing methods in the literature of traditional homogeneous two-sample capture-recapture data (i.e. DRS) when the two lists are thought to be behaviorally dependent.

In the next section, we discuss data structure for DRS and relevant model \emph{$M_{tb}$}. In section \ref{Integrated_Likli_Section}, at first we discuss the integrated likelihood method using weight function as uniform and Jeffrey's densities. Later we propose a novel integrated likelihood method through the construction of unrelated nuisance parameter and informative priors. Evaluation of our proposed method by comparing with some other available Bayesian methods is carried out by an extensive simulation study in section \ref{Numeric-Illustra}. Thereafter, we illustrate our method by applying them to two real life data sets. Finally in section \ref{conclu}, we summarize our findings and provide some comments about the usefulness of the proposed integrated likelihood.

\section{ANALYSES ON DUAL-RECORD SYSTEM: PRELIMINARIES}
\subsection{Dual-record data structure}
The idea of Dual collection came from very popular capture-recapture analysis in wildlife management to estimate population size. Let us consider a human population \emph{U} of size \emph{N} is to be estimated. Any attempt to enlist all the individuals in $U$ is believed to be incomplete as it fails to capture all individuals in that population. In this paper we consider two common basic assumptions that (1) population is closed within the time of two sources gathering information, (2) individuals are homogeneous with respect to capture probabilities. To estimate the true \emph{N}, minimum two sources of information covering that population is needed. When information is collected by two sources and classify all the captured individuals in \emph{U} according to a multinomial fashion (\textit{see} Table \ref{Tab:1}), then it is known as Dual-record System or Dual Collection. This type of classification is obtained by matching the individuals captured by the first (list 1) and second sources (list 2). The total number of distinct captured individuals by the two lists is $x_0$ (say), then $x_0=x_{1\cdot}+x_{\cdot1}-x_{11}$. Clearly, the number of missed individuals $x_{00}$ by both systems is unknown and that makes the total population size \emph{N}($=x_{\cdot\cdot}$) unknown. Expected Proportions or probabilities for each cell are also given in Table \ref{Tab:1} and these notations will be followed throughout this paper.
		\begin{table}[ht]
	\centering
	\caption{$2\times2$ data structure from Dual-record-System (DRS) with corresponding cell probabilities mentioned in [ ] and $p_{\cdot\cdot}$=1.}
	\begin{tabular}{lccc}
		&\multicolumn{3}{c}{List 2} \\
		\cline{2-4}
		List 1 & In & out & Total\\
		\hline \hline
		In & $x_{11}[p_{11}]$ & $x_{10}[p_{10}]$ & $x_{1\cdot}[p_{1\cdot}]$\\
		Out& $x_{01}[p_{01}]$ & $x_{00}[p_{00}]$ & $x_{0\cdot}[p_{0\cdot}]$\\ 
		\hline
		Total& $x_{\cdot1}[p_{\cdot1}]$ & $x_{\cdot0}[p_{\cdot0}]$ & $x_{\cdot\cdot}=N[p_{\cdot\cdot}]$\\
		\hline
	\end{tabular}
	\label{Tab:1}
\end{table}

Combining all the information, estimate of \emph{N} could be obtained assuming different conditions on the individual's capture probabilities leading to different models. A very common practice, across all fields of applications, is to assume casual independence between two lists' probabilities. Hence, the conditional likelihood estimate from the resulting multinomial model (denoted as $M_t$) is $\hat{N}_{ind}=[x_{1\cdot}x_{\cdot1}/x_{11}]$, which is popularly known as Lincoln-Petersen estimator (Otis et al, 1978) or dual system estimator (DSE); \textit{see} Wolter (1986) and Chatterjee and Mukherjee (2016a) for details. But this model is highly criticized by several statisticians and practitioners due to the failure of its independence assumption in real life applications. In demographic study, violation of causal independence often is commonly observed due to positive correlation between two sources (or lists) of counts (\textit{see} Chandrasekar and Deming, 1949; Nour, 1982;). Assuming such positive dependency, Nour(1982) proposed an estimate of $N$ as \[\hat{N}_{Nour}=x_0+\frac{2x_{11}x_{10}x_{01}}{(x_{11}^2+x_{10}x_{01})}.\] Negative dependence is observed in case of epidemiological surveillance of rare or critical disease, like HIV, drug abusing, etc. 

\subsection{Model $M_{tb}$}\label{model_Mtb}
Causal independence assumption is criticised in surveys and censuses of human populations. The concern is that an individual's probability of capture in List 1 may be change in response to capture in the second list. An individual who is captured in first attempt may have more (or less) chance to be included in the List 2 than the individual who has not been captured in first attempt. The change may occur due to different causes (\textit{see} Wolter, 1986). This change is grossly known as behavioral response variation. When this chance is more, the corresponding individuals are treated as \emph{recapture prone}, when chance is less, individuals are treated to be \emph{recapture averse}. When this feature is combined with the time variation effect, one would get a complex model denoted as $M_{tb}$. To model this behavioral response variation generally, let us consider the following notations:

Prob(An individual present in List 1) = $p_{1\cdot}$,

Prob(An individual present in List 2 $|$ not present in List 1) = $p_{01}/(1-p_{1\cdot})$ = $p$ and

Prob(An individual is captured in List 2 $|$ captured in List 1) = $p_{11}/p_{1\cdot}$ = $c$.

Therefore, the likelihood function for model $M_{tb}$ in DRS is
	\begin{eqnarray}
	L(N,p_{1\cdot},p,c) &\propto& \frac{N!}{(N-x_0)!}c^{x_{11}}p_{1\cdot}^{x_{1\cdot}}p^{x_{01}}(1-p_{1\cdot})^{N-x_{1\cdot}}(1-p)^{N-x_{0}}(1-c)^{x_{10}},\label{Eq_L_1}
	\end{eqnarray}
for $N>x_0$, $0<p_{1\cdot}, p, c<1$, consists lesser number of sufficient statistics ($x_{11},x_{01},x_{10}$) than the parameters ($N,p_{1\cdot},p,c$) (\textit{see} Otis et al., 1978). One can consider a popular assumption that recapture probability at second sample, \emph{c}, is equal to a constant multiple of the conditional probability $p$, hence, $c=\phi p$. Chao et al. (2000) adopted this relation from Lloyd (1994) to get rid of the problem. Then reparameterized version of likelihood (\ref{Eq_L_1}) becomes
\begin{center}
	\begin{eqnarray}
	L_{tb}(N,p_{1\cdot},p,\phi) &\propto& \frac{N!}{(N-x_0)!}\phi^{x_{11}}p_{1\cdot}^{x_{1\cdot}}p^{x_{\cdot1}}(1-p_{1\cdot})^{N-x_{1\cdot}}(1-p)^{N-x_{0}}(1-\phi p)^{x_{10}}\label{Eq_L_2}
	\end{eqnarray}
\end{center}
and $\phi$ is termed as the \textit{behavioral response effect} characterizing the behavioral dependency of an individual belongs to the population at the time of second capture. In Equation (\ref{Eq_L_2}), $\phi$ is orthogonal to $N$. It can be noticed that, in DRS, the dimension cannot be reduced by any reparametrization and therefore, identifiability problem persists in Equation (\ref{Eq_L_2}) also. $\phi$ and $p$ are not estimable separately but their product $c$ is rather estimable. However, this second form of parameterization in Equation (\ref{Eq_L_2}) may be of interest in lieu of Equation (\ref{Eq_L_1}) (\textit{see} Chao et al., 2000) as Equation (\ref{Eq_L_2}) is characterized by the parameter $\phi$, which has a clear implication to define the nature of underlying behavioral dependence among two sources, i.e. whether the given population is \textit{recapture prone} or \textit{averse}. Replacement of $p$ with $c/\phi$ in Equation (\ref{Eq_L_1}) is another version of parametrization and in this form, $\phi$ is not orthogonal to $N$.

\section{INTEGRATED LIKELIHOOD METHOD}\label{Integrated_Likli_Section}
Let us consider a statistical model with likelihood function $L(\lambda|\underline{\textbf{x}})$ with $\lambda=(\theta, \psi)$, where $\theta (\in \Theta)$ is parameter of interest and $\psi (\in \Psi)$ represents nuisance parameter. Both the $\theta$ and $\psi$ may be vector valued. Presence of more nuisance parameters in the model affects the comparative inferential study based on the likelihood (\textit{see} Severini, 2000). Now our aim is to find a function that can summarize the set of likelihoods $\mathcal{L}^{*}=\{L(\theta,\psi|\underline{\textbf{x}}):\psi\in\Psi\}$ over $\Psi$. That summerized function $L^*(\theta)$ of $\theta$, is some extent used as if the inference frame has $\theta$ as the full parameter and therefore has likelihood function, $L^*(\theta)$. We refer such functions $L^*(\theta)$ here as pseudo likelihood function of $\theta$. Construction of such pseudo-likelihood is performed by elimination nuisance parameter which can be handled by \textit{integrated likelihood method}. In this approach nuisance parameter is eliminated through integration or rather it can be said that the set of likelihoods $\mathcal{L}^{*}=\{L(\theta,\psi|\underline{\textbf{x}}):\psi\in\Psi\}$ is summarized over $\Psi$ by a weighted average with respect to a chosen function on $\Psi$, say, $\pi(\psi|\theta)$ defined on $\Psi$. Hence, integrated likelihood function with respect to the weight $\pi(\psi|\theta)$ is
\begin{eqnarray}
L^{I}(\theta)&=&\int_{\Psi}{L(\theta,\psi|\textbf{\underline{x}})\pi(\psi|\theta)d\psi};\label{Eq_IMLE_0}
\end{eqnarray}
\textit{see} Severini (2000) for detailed discussion. One advantage of integrated likelihood over pseudo-likelihoods (conditional, marginal) is that it is always possible to construct unlike conditional or marginal likelihood. However, it is not necessary to choose $\pi(\psi|\theta)$ as a proper density function in this context. But one drawback includes the plausible subjectiveness due to $\pi(\psi|\theta)$. The basic aim always remains to choose a suitable $\pi(\psi|\theta)$ such that $L^{I}(\theta)$ could be efficiently useful for non-Bayesian likelihood inference. The \textit{mle} of $L^{I}(\theta)$ is to be treated as the resulting estimate of $N$ from this method.

In capture-recapture context, for fixed $\theta$, Jeffrey's and uniform priors are the two most popular non-informative prior densities on $\psi$ (Salasar, 2014). When $\theta=N$, $\psi=(p_{1\cdot},c,p)$ and uniform prior $\pi(\psi|N)\propto1$ is chosen for $\psi$, then from Equation (\ref{Eq_IMLE_0}), integrate likelihood from likelihood (\ref{Eq_L_1}) becomes
\[L_{U}^{I}(N)=\int_{\Psi}{L(\theta,\psi|\textbf{\underline{x}})}\propto(N+1)^{-1}(N-x_{1\cdot}+1)^{-1},\]
for $N>x_0$. Clearly, $L_{U}^{I}(N)$ is strictly decreasing over its domain. Hence, $L_{U}^{I}(N)$ fails. Again, if we consider $\pi(\psi|N)$ as Jeffrey's prior, then $\pi(\psi|\theta)\propto\sqrt{|\mathcal{I}_{N}(\psi)|}$, where $\mathcal{I}_{N}(\psi)$ is $3\times3$ Fisher's information matrix. Therefore,
\begin{eqnarray}
\pi(\psi|N)&\propto&\left[det\left\{Diag\left(\frac{N}{p_{1\cdot}(1-p_{1\cdot})}, \frac{Np_{1\cdot}}{c(1-c)}, \frac{N(1-p_{1\cdot})}{p(1-p)}\right)\right\}\right]^{1/2}\nonumber\\
&=&\{c(1-c)p(1-p)\}^{-1}\label{Wt_1}.
\end{eqnarray}
Hence, using the weight in Equation (\ref{Wt_1}) in Equation (\ref{Eq_IMLE_0}), one would have the integrate likelihood corresponds to likelihood (\ref{Eq_L_1}) as
\begin{eqnarray}
L_{J}^{I}(N)&=& \int_{\Psi}{L(\theta,\psi|\textbf{\underline{x}})\pi(\psi|\theta)d\psi}=\frac{(N-x_{1\cdot})}{(N+1)(N-x_{0})},\nonumber
\end{eqnarray}
which is of $O(N^{-1})$ and hence, strictly decreasing over its domain $N>x_0$. Note that both of the above integrated likelihoods based on non-informative priors fails, because these priors could not add subjective information for $\psi$ in the likelihood (\ref{Eq_L_1}) so that it became well-behaved and produce reasonable estimate for $N$.

\subsection{Proposed Integrated Likelihood approach}\label{Sec:Pro_Int_lik_Meth}
In order to construct an integrated likelihood function to be useful for the present likelihood model, suitable informative prior should be used. There are two alternative parameterizations Equations (\ref{Eq_L_1}) and (\ref{Eq_L_2}) of the model $M_{tb}$ in section \ref{model_Mtb} and we develope our theoretical findings for both parameterizations on different $\psi$. To begin, let us start by fetching the idea of strongly unrelated parameters defined in Severini (2007). 
\begin{defn}
	Two parameters $\gamma$ and $\theta$ are said to be \textit{strongly unrelated} if 
	\[\hat{\gamma}_{\theta}=\hat{\gamma}+O(N^{-1/2})O(|\theta-\hat{\theta}|)  \]
	holds.
\end{defn}
Henceforth, we consider $\gamma$ as a nuisance parameter with the same dimension as $\psi$ and $\gamma$ is \textit{strongly unrelated} to $\theta$, the interest parameter here. Therefore, we follow the prior choosing mechanism developed in Severini (2007). This mechanism helps us to find a \textit{strongly unrelated} nuisance parameter $\gamma$, in terms of $\psi$ and $\theta$ for model $M_{tb}$, in such a way so that $\gamma$ and $\theta$ would become independent under $\pi(\psi|\theta)$, i.e., $\pi(\psi|\theta)=\pi(h(\gamma,\theta)|\theta)=\pi(\gamma|\theta)=\pi(\gamma)$. Therefore, the task is to find such parameter $\gamma$ and then choose a prior density $\pi(\gamma)$ for $\gamma$ that does not depend on $\theta$. Hence, the integrated likelihood function for $\theta$ with respect to $\pi(\gamma)$ is
\begin{eqnarray}\label{Eq_05}
\overline{L}^{I}(\theta)&=&\int_{\Gamma}{L(\theta,\gamma|\textbf{\underline{x}})\pi(\gamma)d\gamma}.
\end{eqnarray}
Construction of such nuisance parameter $\gamma$ is discussed below.

We consider the following equation from Severini(2007):
\begin{eqnarray}
E\{\ell_{\psi}(\theta,\psi);\hat{\theta},\gamma\}&\equiv & E\{\ell_{\psi}(\theta,\psi)
;\theta_0,\gamma_0\}\mid_{(\theta_0=\hat{\theta},\gamma_0=\gamma)}=0,\label{Eq_IMLE_1}
\end{eqnarray}
from which one can solve $\gamma$ as $\gamma(\theta,\psi;\hat{\theta})$. Severini (2007) proved that $\hat{\gamma}=\hat{\psi}$ and $\gamma$ is \textit{strongly unrelated} to $\theta$ i.e. $\hat{\gamma}_{\theta}=\hat{\gamma}+O(n^{-1/2})O(|\theta-\hat{\theta}|)$. Then solution $\gamma(\theta,\psi;\hat{\theta})$ is called zero-score-expectation parameter. Now, one can choose any suitable prior $\pi(\gamma)$ for $\gamma$ as $\overline{L}^{I}(\theta)$ in Equation (\ref{Eq_05}) does not heavily depend on the chosen prior whereas for orthogonal parameters, proposed integrated likelihood (\ref{Eq_05}) may be dependent to the choice of prior.

One can find $\gamma$ in different way. The aim is to find a function $\gamma(\theta,\psi)$ such that $\hat{\gamma}_{\theta}=\hat{\gamma}+O(n^{-1/2})O(|\theta-\hat{\theta}|)$. Hence, one can find such a parameter $\gamma=g(\theta,\psi)$, which implies $\psi=h(\theta,\gamma)$ for some \textit{h}, if exists so that $\hat{\psi}_{\theta}=h(\theta,\hat{\gamma}_{\theta})=h(\theta,\hat{\gamma})+O(n^{-1/2})$. So, for any value of $\theta$, $\hat{\psi}_{\theta}$ depends on the data only through $\hat{\gamma}$. In many situations, this $\hat{\gamma}$ does not exist. So, we consider $\gamma$ as a function of $\hat{\theta}$ in addition to ($\theta,\psi$). It can be written as $\gamma=g(\theta,\psi;\hat{\theta})$ which implies $\psi=h(\theta,\gamma;\hat{\theta})$. As $\hat{\psi}_{\theta}=h(\theta,\hat{\gamma}_{\theta};\hat{\theta})$, then we must have
\begin{center}
	\begin{eqnarray}
	\hat{\psi}_{\theta}&=&h(\theta,\hat{\gamma};\hat{\theta})+O(n^{-1/2})O(|\theta-\hat{\theta}|)\label{Eq_IMLE_2},
	\end{eqnarray}
\end{center}
when $\hat{\gamma}_{\theta}=\hat{\gamma}+O(n^{-1/2})O(|\theta-\hat{\theta}|)$. Hence, one wish to find such a function $h(\cdot)$ so that Equation (\ref{Eq_IMLE_2}) holds.

Let us address some desirable properties of $\overline{L}^{I}(\theta)$ in the context of model $M_{tb}$. By construction, unrelated parameter $\gamma$ is less related to $\theta$ than any nuisance parameter. Hence, \textit{prior sensitiveness} to $\theta$ can be reduced by the use of $\pi(\gamma)$ in lieu of $\pi(\psi|\theta)$. Moreover, \textit{score unbiasedness} and \textit{information unbiasedness} criteria are incorporated in the present construction and the resulting likelihood estimate is invariant with respect to the different parameterizations. For detailed discussion of these properties, we refer Severini (2007).

Now we discuss the application of the proposed idea of integrated likelihood method to model $M_{tb}$ in the context of DRS. We construct the relevant unrelated nuisance parameter $\gamma$, then choose $\pi(\gamma)$ satisfying posterior unbiasedness condition. Firstly, we present the consequent theorems, results and properties for the parametrization in likelihood (\ref{Eq_L_1}). Further, we also analyse the parametrization in likelihood (\ref{Eq_L_2}) following similar course of theorems and results as for parametrization in likelihood (\ref{Eq_L_1}).

\subsubsection*{\textbf{\textit{Implementation for the likelihood (\ref{Eq_L_1})}}}\label{Imple_Lik01}
Let us consider the parameter of interest $\theta=N$. Following theorem finds strongly unrelated parameter corresponding to nuisance parameter $\psi=(p_{1\cdot},c,p)$. Proof is given in the \textit{Appendix}.
%
\begin{theorem}{Theorem 3.1.}{}
	For parametrization in (\ref{Eq_L_1}), consider $\theta=N$ and $\psi=(p_{1\cdot},c,p)$. Then using Equation (\ref{Eq_IMLE_1}), strongly unrelated parameter is \textbf{$\gamma$}=($\gamma_1,\gamma_2,\gamma_3$), where $\gamma_1=(N/\hat{N}_{ind})p_{1\cdot}$, $\gamma_2=c$ and $\gamma_3=\frac{p(1-p_{1\cdot})}{(\hat{N}_{ind}/N)-p_{1\cdot}}$.
\end{theorem}\label{Theo_1}

Since current model suffers from non-identifiability, so non-informative priors for resultant unrelated parameters $\gamma$ (in Theorem \ref{Theo_1}) would not work satisfactorily, as in the case of ordinary integrated likelihoods under uniform and Jeffrey's priors. Thus, we consider some informative prior for $\gamma$ subject to the condition that hyper-parameters satisfy some relations that lead to a integrated likelihood.
\begin{result}\label{Res_1}
	In connection with Theorem \ref{Theo_1}, if the prior $\pi(\gamma)$ is of the form $\pi(\gamma)=\pi(\gamma_1)\pi(\gamma_2)\pi(\gamma_3)$ and $\pi(\gamma_1)=GB1(b_1=\frac{N}{\hat{N}_{ind}},r_1,s_1)$, $\pi(\gamma_2)=Unif(0,1)$ and $\pi(\gamma_3)=GB1(b_2=\frac{1-p_{1\cdot}}{(\hat{N}_{ind}/N)-p_{1\cdot}},r_2,s_2)$, for any positive real numbers $r_2$, $s_2$, $r_1$, $s_1$ satisfying $r_2+s_2=s_1$, where GB1() stands for \textit{Generalized Beta distribution of Type 1}, then using Equation (\ref{Eq_05}) integrated likelihood of $N$ for model $M_{tb}$ becomes
	\begin{eqnarray}
	\overline{L}_{tb}^{I}(N) &=& \frac{\Gamma(N-x_0+s_2)\Gamma(N+1)}{\Gamma(N+r_1+s_1)\Gamma(N-x_0+1)},\label{Eq_IMLE_3}
	\end{eqnarray}
	for $N\geq x_0$, where $\Gamma(a)$ denotes the Gammaa function of $a$ equivalent to $\int_{0}^{\infty}e^{-x}x^{a-1}$. Thus, for the given values of $r_1$, $s_1$ and $s_2$, $\overline{L}_{tb}^{I}(N)$, in Equation (\ref{Eq_IMLE_3}), is non-decreasing in $N$ for
	\begin{center}
		\begin{eqnarray}
		N\leq\frac{x_{0}(r_1+s_1-1)}{r_2+r_1}-1\label{Eq_09}
		\end{eqnarray}
	\end{center}
	and $\overline{L}_{tb}^{I}(N)$ converges to 0 as $N\rightarrow\infty$.
\end{result}
\textit{Proof.} Given in the \textit{Appendix}.

The rationale behind the assumption $r_2+s_2=s_1$ primarily is to make the resulting integrate likelihood well-behaved, and secondarily, to reduce dimension of the hyper-parameters in $\pi(\gamma)$. Now, on the basis of the relationship between $\gamma$ and $\psi$, mentioned in Theorem \ref{Theo_1}, we suggest the values of hyper-parameters $r_2$, $s_2$ and $r_1$ using posterior unbiasedness of $\gamma$ (after considering $\gamma$ as function of $N$). We suggest $r_2=bx_{12}$ and $s_2=b(N^*-x_0)$ for some real positive $b$, where $N^*$ is some other working estimate of $N$. So, $s_1=b(N^*-x_{1\cdot})$. Now, for fixed $N$, $\gamma_1=(N/\hat{N}_{ind})p_{1\cdot}=(N/\hat{N}_{ind})(x_{1\cdot}/N)=(x_{1\cdot}/\hat{N}_{ind})$. Hence, from the posterior unbiasedness condition regarding unrelated parameter $\gamma_1$, we have $(x_{1\cdot}/\hat{N}_{ind})\simeq E_{\pi}(\gamma_1)=(N/\hat{N}_{ind})\{(r_1+x_{1\cdot})/(r_1+s_1+N)\}$ and this implies $r_1=x_{1\cdot}s_1/(N-x_{1\cdot})=r_1(N, s_1)$, say, after some algebraic manipulation.
Thus, $r_1$ depends on $N, s_1$ and therefore, the right hand side of the condition in Equation (\ref{Eq_09}) also becomes dependent on $N$. So, $N\leq \frac{x_{0}\{r_1(N)+s_1-1\}}{\{r_2+r_1(N)\}}-1\Leftrightarrow\frac{x_{0}\{r_1(N)+s_1-1\}}{\{r_2+r_1(N)\}}-N\geq1$ and $\frac{x_{0}\{r_1(N)+s_1-1\}}{\{r_2+r_1(N)\}}-N$ is non-increasing in $N$ for fixed $s_1$ and $r_2$. Hence, the result stated in the following theorem, obtained from a very easy algebraic simplification, discusses the possibility of existence of the corresponding maximum likelihood estimate. Proof is sketched in the \textit{Appendix}.
\begin{theorem}{}{}\label{Theo_2}
	For $r_1=x_{1\cdot}s_1/(N-x_{1\cdot})=r_1(N, s_1)$, $\exists$ a real number, say $N_0=N_0(\underline{\textbf{x}},s_1,r_2)$, for which inequality in Equation (\ref{Eq_09}) is equivalent to $N\leq N_{0}<\infty$.
\end{theorem}
Hence from Theorem \ref{Theo_2}, it can be said that $\overline{L}_{tb}^{I}(N)$ is increasing in $N$ for $N\leq N_{0}$ and hence, estimate based on our proposed integrated lkelihood method is $\hat{N}_{tb}^{I}=[N_0]+1$, if $N_0$ is not an integer and $\hat{N}_{tb}^{I}=[N_0]$ and $[N_0]+1$ if $N_0$ is an integer. The expression for $N_0$ turns out to be mathematically intractable and thus obtaining an explicit solution is not possible. We explore several properties of the resulting estimator, as pointed out in Severini(2007), through computation in section \ref{Simulation}. To implement the above prior specification, we suggest $N^*=(\hat{N}_{Nour}+\hat{N}_{ind})/2$ and $b=(1+(\hat{N}_{ind}-((x_0+\hat{N}_{ind})/2)-1)^{-1})/2$, when we don't know anything about the plausible direction of $\phi$. On the other hand, if we know that $\phi>1$ (recapture proneness), then $N^*=\hat{N}_{Nour}$ and $b=1$. If $\phi<1$ (recapture aversion), suggested $N^*=\hat{N}_{ind}$ and $b=(\hat{N}_{ind}-((x_0+\hat{N}_{ind})/2)-1)^{-1}$.

\begin{rem}
If anyone is interested to consider $\theta=(N,c,p)$, where $c=\phi p$, relevant for likelihood (\ref{Eq_L_1}) then strongly unrelated parameters can be obtained from the relevant log-likelihood functions as before. Now, for $\psi=p_{1\cdot}$, if the prior is taken on the associated unrelated parameter $\gamma$ as $\pi(\gamma)=GB1(b=\frac{N}{\hat{N}_{ind}},r,s)$, for any positive real numbers $r$ and $s$; then integrated likelihood for $N(\geq x_0)$ reduces to
	\begin{center}
		\begin{eqnarray}
		\overline{L}_{M_{tb}}^{I}(N,c,p) &=& \frac{\Gamma(N-x_{1\cdot}+s)\Gamma(N+1)}{\Gamma(N+r+s)\Gamma(N-x_0+1)}c^{x_{11}}p^{x_{01}}(1-c)^{x_{10}}(1-p)^{N-x_{0}},\nonumber
		\end{eqnarray}
	\end{center}
	and it fails to produce the \textit{mle} of $\theta=(N,c,p)$. Perhaps the failure is due to the scarcity of enough information to make inference about the two parameters $N$ and $\phi$, which are actually orthogonal to each other. For details about parameter orthogonality, \textit{see} Cox and Reid (1987).
\end{rem}

\section{NUMERICAL ILLUSTRATION}\label{Numeric-Illustra}
\subsection{Simulation study}\label{Simulation}
In this section we conduct a simulation study to evaluate the performance of our proposed approach and understand its efficiency in applying the method using the available directional knowledge on $\phi$. This study is designed as follows. Let us simulate hypothetical populations corresponding to six pairs of capture probabilities ($p_{1\cdot}, p_{\cdot1}$)=\{(0.50, 0.65), (0.60, 0.70), (0.80, 0.70), (0.70, 0.55), (0.55, 0.75), (0.70, 0.50)\} for each case of \emph{recapture prone} (represented by $\phi=1.25, 1.50$) and \emph{recapture averse} (represented by $\phi=0.60, 0.80$) situations. We denote six populations corresponding to the six pairs of ($p_{1\cdot}, p_{\cdot1}$) as P1, P2, P3, P4, P5 and P6 respectively for \emph{recapture prone} situations and associated results are presented in Tables \ref{Tab:2a} and \ref{Tab:2b} for true population size $N=200$ and $500$ respectively. Results of the other six populations comprising the same six pairs of ($p_{1\cdot}, p_{\cdot1}$) reflecting \emph{recapture averse} situations, namely A1-A6 are shown in Tables \ref{Tab:3a} and \ref{Tab:3b} for the same two $N$ values respectively. 1000 data sets on ($x_{1\cdot},x_{\cdot1},x_{11}$) are generated from each of the 12 populations. Our proposed integrated likelihood estimate have been obtained for each data set. Finally, estimate of $\hat{N}_{tb}^{I}$ is obtained by averaging over 1000 posterior means. Based on those 1000 estimates, the sample RMSE (Root Mean Square Error) and $95\%$ confidence interval (C.I.) are computed. In addition to that, we also compute similar statistics for \textit{Lee}'s and \textit{SEMWiG} method, proposed respectively in Lee et al. (2003) and Chatterjee and Mukherjee (2016b), to compare the performance of our newly proposed non-Bayesian method. However, Lee et al. (2003) illustrated their approach in the context of animal capture-recapture experiment with a large number of sampling occasions. To compute the estimates using \textit{Lee} and \textit{SEMWiG} methods in this article, we use same priors as considered in Chatterjee and Mukherjee (2016b) when information is available on the directional nature of $\phi$.  Details of their computation strategy, particularly for DRS, can be found in Chatterjee and Mukherjee (2016b). All of these comparison results are summarized in Tables \ref{Tab:2a}-\ref{Tab:3b}.

\begin{table}[h]
	\tiny
	\caption{Summary results of the three methods (Lee, SEMWiG \& $\hat{N}_{tb}^{I}$) applied to the populations P1-P6 with $N=200$ and when $\phi>1$ is known.}
	\begin{minipage}{12cm}
		\begin{tabular}{|lrcccccc|}
			\hline
			Method & & P1 & P2 & P3 & P4 & P5 & P6\\
			\hline
			\multicolumn{8}{|c|}{$\phi=1.25$}\\
			$Lee$\footnote{Prior on $\phi$ is chosen as \textit{U}($1, 2$) since $\phi>1$ is known (\textit{see} Chatterjee and Mukherjee, 2016b)}  & $\hat{N}$(RMSE) & 193(15.38) & 191(12.33) & 195(6.90) & 199(9.71) & 188(14.05) & 204(14.56) \\
			&  C.I.  & $(171, 226)$ & $(176, 212)$ & $(187, 208)$ & $(179, 227)$ & $(175, 208)$ & $(179, 237)$\\
			$SEMWiG$\footnote{Prior on $\phi$ is chosen as \textit{U}($1,p^{-1}$) since $\phi>1$ is known (\textit{see} Chatterjee and Mukherjee, 2016b)}  & $\hat{N}$(RMSE) & 197(12.79) & 194(10.59) & 198(5.89) & 204(10.50) & 190(10.38) & 225(22.89)\\
			&  C.I.  & $(194, 201)$ & $(192, 197)$ & $(197, 200)$ & $(200, 208)$ & $(189, 191)$ & $(220, 230)$ \\
			$\hat{N}_{tb}^{I}$  & $\hat{N}$(RMSE) & 191(13.16) & 195(9.22) & 199(5.07) & 200(6.69) & 190(12.98) & 198(8.99)\\
			& C.I.  & $(173, 208)$ & $(181, 209)$ & $(188, 211)$ & $(186, 215)$ &$(174, 202)$& $(183, 215)$ \\
			&  &  &  &  && & \\
			\multicolumn{8}{|c|}{$\phi=1.50$}\\
			$Lee^a$  & $\hat{N}$(RMSE) & 173(28.33) & 178(23.11) & 189(11.48) & 187(15.64) & 174(26.09) & 190(14.91)\\
			& C.I.  & $(159, 196)$ & $(167, 194)$ & $(183, 201)$ & $(172, 209)$ &$(166, 189)$& $(172, 216)$ \\
			$SEMWiG^b$  & $\hat{N}$(RMSE) & 176(25.80) & 179(21.94) & 191(10.53) & 192(12.40) & 173(26.98) & 194(12.41)\\
			&  C.I.  & $(174, 178)$ & $(178, 181)$ & $(189, 193)$ & $(190, 194)$ & $(172, 175)$ & $(192, 197)$ \\
			$\hat{N}_{tb}^{I}$  & $\hat{N}$(RMSE) & 175(26.99) & 180(21.66) & 191(10.56) & 191(11.85) & 175(26.54) & 192(11.00)\\
			&  C.I.  & $(157, 192)$ & $(166, 195)$ & $(181, 202)$ & $(177, 208)$ & $(160, 188)$ & $(177, 210)$\\
			\hline
		\end{tabular}
		\label{Tab:2a}
	\end{minipage}
\end{table}

\begin{table}[h]
	\tiny
	\caption{Summary results of the three methods (Lee, SEMWiG \& $\hat{N}_{tb}^{I}$) applied to the populations P1-P6 with $N=500$ and when $\phi>1$ is known.}
	\begin{minipage}{12cm}
		\begin{tabular}{|lrcccccc|}
			\hline
			Method & & P1 & P2 & P3 & P4 & P5 & P6\\
			\hline
			\multicolumn{8}{|c|}{$\phi=1.25$}\\
			$Lee$\footnote{Prior on $\phi$ is chosen as \textit{U}($1, 2$) since $\phi>1$ is known (\textit{see} Chatterjee and Mukherjee, 2016b)}  & $\hat{N}$(RMSE) & 472(34.11) & 478(24.33) & 490(12.02) & 488(17.03) & 504(13.71)& 527(32.80)\\
			& C.I.  & $(438, 518)$ & $(451, 519)$ & $(473, 515)$ & $(457, 531)$ &$(445, 576)$& $(471, 599)$ \\
			$SEMWiG$\footnote{Prior on $\phi$ is chosen as \textit{U}($1,p^{-1}$) since $\phi>1$ is known (\textit{see} Chatterjee and Mukherjee, 2016b)}  & $\hat{N}$(RMSE) & 484(23.92) & 485(21.25) & 495(9.87) & 502(16.32) & 480(24.64) & 514(22.89)\\
			&  C.I.  & $(451, 522)$ & $(456, 511)$ & $(477, 512)$ & $(474, 532)$ & $(473, 493)$ & $(505, 521)$ \\
			$\hat{N}_{tb}^{I}$  & $\hat{N}$(RMSE) & 478(26.75) & 486(19.60) & 498(9.06) & 497(13.85) & 476(27.19) & 496(13.35)\\
			&  C.I.  & $(451, 508)$ & $(460, 511)$ & $(480, 514)$ & $(471, 521)$ & $(451, 500)$ & $(471, 519)$\\
						&  &  &  &  && & \\
			\multicolumn{8}{|c|}{$\phi=1.50$}\\
			$Lee^a$  & $\hat{N}$(RMSE) & 437(64.17) & 455(47.37) & 481(21.58) & 483(28.67) & 44556.40 & 492(20.01)\\
			& C.I.  & $(405, 487)$ & $(424, 500)$ & $(452, 522)$ & $(437, 538)$ &$(419, 488)$& $(449, 550)$ \\
			$SEMWiG^b$  & $\hat{N}$(RMSE) & 442(59.56) & 450(51.53) & 477(22.66) & 476(27.53) & 436(64.69) & 480(25.67)\\
			&  C.I.  & $(436, 452)$ & $(447, 456)$ & $(475, 582)$ & $(470, 482)$ & $(433, 439)$ & $(475, 487)$\\
			$\hat{N}_{tb}^{I}$  & $\hat{N}$(RMSE) & 442(60.52) & 455(47.24) & 480(21.69) & 479(23.62) & 439(62.40) & 479(24.25)\\
			&  C.I.  & $(415, 469)$ & $(432, 478)$ & $(464, 496)$ & $(455, 501)$ & $(415, 463)$ & $(452, 503)$\\
			\hline
		\end{tabular}
		\label{Tab:2b}
	\end{minipage}
\end{table}

\begin{table}[h]
	\tiny
	\caption{Summary results of the three methods (Lee, SEMWiG \& $\hat{N}_{tb}^{I}$) applied to the populations A1-A6 with $N=200$ and when $\phi<1$ is known.}
	\begin{minipage}{12cm}
		\begin{tabular}{|lrcccccc|}
			\hline
			Method & & A1 & A2 & A3 & A4 & A5 & A6\\
			\hline
			\multicolumn{8}{|c|}{$\phi=0.60$}\\
			$Lee$\footnote{Prior on $\phi$ is chosen as \textit{U}($0.2, 1.4$) since $\phi<1$ is known (\textit{see} Chatterjee and Mukherjee, 2016b)} & $\hat{N}$(RMSE)  & 236(43.96) & 222(28.77) & 214(15.51) & 222(26.56) & 234(38.12) & 224(29.91) \\
			& C.I.  & $(186, 316)$ & $(195, 271)$ & $(200, 244)$ & $(188, 291)$ & $(198, 295)$ & $(186, 297)$\\
			$SEMWiG$\footnote{Prior on $\phi$ is chosen as \textit{U}($\hat{c}, 1$) since $\phi<1$ is known (\textit{see} Chatterjee and Mukherjee, 2016b)}  & $\hat{N}$(RMSE) & 213(13.44) & 211(11.28) & 208(7.90) & 200(\textbf{0.76}) & 216(16.71) & 199(2.05)\\
			&  C.I.  & $(200, 230)$ & $(203, 219)$ & $(203, 211)$ & $(193, 209)$ & $(210, 222)$ & $(193, 204)$\\
			$\hat{N}_{tb}^{I}$  & $\hat{N}$(RMSE) & 214(15.72) & 215(16.23) & 210(11.26) & 203(5.42) & 220(21.52) & 200(4.35)\\
			&  C.I.  & $(197, 230)$ & $(205, 226)$ & $(206, 215)$ & $(194, 215)$ & $(211, 231)$ & $(190, 214)$\\
			\multicolumn{8}{|c|}{}\\
			\multicolumn{8}{|c|}{$\phi=0.80$}\\
			$Lee^a$  & $\hat{N}$(RMSE) & 196(7.78) & 194(8.64) & 199(3.89) & 191(10.92) & 199(9.26) & 203(11.26)\\
			&  C.I.  & $(169, 242)$ & $(184, 220)$ & $(193, 213)$ & $(178, 247)$ & $(186, 229)$ & $(175, 258)$\\
			$SEMWiG^b$  & $\hat{N}$(RMSE) & 192(8.43) & 195(4.70) & 195(4.48) & 192(9.34) & 199(3.45) & 185(14.62)\\
			&  C.I.  & $(188, 198)$ & $(189, 200)$ & $(193, 197)$ & $(186, 196)$ & $(194, 204)$ & $(182, 190)$\\
			$\hat{N}_{tb}^{I}$  & $\hat{N}$(RMSE) & 196(8.25) & 201(5.32) & 201(3.25) & 193(9.32) & 201(4.39) & 190(11.80)\\
			& C.I.  & $(181, 208)$ & $(190, 212)$ & $(195, 207)$ & $(183, 206)$ & $(193, 215)$ & $(178, 202)$\\
			\hline
		\end{tabular}
	\end{minipage}
	\label{Tab:3a}
\end{table}

\begin{table}[h]
	\tiny
	\caption{Summary results of the three methods (Lee, SEMWiG \& $\hat{N}_{tb}^{I}$) applied to the populations A1-A6 with $N=500$ and when $\phi<1$ is known.}
	\begin{minipage}{12cm}
		\begin{tabular}{|lrcccccc|}
			\hline
			Method & & A1 & A2 & A3 & A4 & A5 & A6\\
			\hline
			\multicolumn{8}{|c|}{$\phi=0.60$}\\
			$Lee$\footnote{Prior on $\phi$ is chosen as \textit{U}($0.2, 1.4$) since $\phi<1$ is known (\textit{see} Chatterjee and Mukherjee, 2016b)}  & $\hat{N}$(RMSE) & 625(136.80) & 608(115.19) & 553(56.30) & 530(34.31) & 619(124.43) & 529(40.26)\\
			& C.I.  & $(485, 827)$ & $(505, 744)$ & $(511, 619)$ & $(472, 671)$ & $(510, 775)$ & $(464, 669)$\\
			$SEMWiG$\footnote{Prior on $\phi$ is chosen as \textit{U}($\hat{c}, 1$) since $\phi<1$ is known (\textit{see} Chatterjee and Mukherjee, 2016b)}  & $\hat{N}$(RMSE) & 521(25.48) & 517(18.23) & 517(17.16) & 517(19.69) & 527(27.73) & 513(15.87)\\
			&  C.I.  & $(490, 549)$ & $(505, 529)$ & $(514, 521)$ & $(499, 539)$ & $(512, 542)$ & $(491, 538)$\\
			$\hat{N}_{tb}^{I}$  & $\hat{N}$(RMSE) & 541(43.00) & 534(35.24) & 528(28.34) & 514(16.47) & 548(49.80) & 505(10.61)\\
			&  C.I.  & $(519, 570)$ & $(520, 550)$ & $(522, 535)$ & $(500, 528)$ & $(535, 566)$ & $(486, 520)$\\
			\multicolumn{8}{|c|}{}\\
			\multicolumn{8}{|c|}{$\phi=0.80$}\\
			$Lee^a$  & $\hat{N}$(RMSE) & 472(35.42) & 490(13.65) & 509(10.34) & 481(20.08) & 531(39.79) & 529(38.53)\\
			& C.I.  & $(431, 565)$ & $(460, 560)$ & $(485, 545)$ & $(448, 545)$ & $(475, 633)$ & $(459, 608)$\\
			$SEMWiG^b$  & $\hat{N}$(RMSE) & 478(25.27) & 492(11.74) & 498(5.28) & 484(19.34) & 490(12.15) & 480(22.96)\\
			&  C.I.  & $(455, 507)$ & $(477, 510)$ & $(489, 508)$ & $(465, 506)$ & $(485, 496)$ & $(462, 493)$\\
			$\hat{N}_{tb}^{I}$  & $\hat{N}$(RMSE) & 487(17.81) & 499(8.30) & 502(5.29) & 484(17.55) & 506(10.19) & 488(13.58)\\
			&  C.I.  & $(465, 512)$ & $(484, 515)$ & $(492, 511)$ & $(470, 498)$ & $(490, 521)$ & $(469, 505)$\\
			\hline
		\end{tabular}
	\end{minipage}
	\label{Tab:3b}
\end{table}

From the Tables \ref{Tab:2a}-\ref{Tab:3b} it can be clearly noticed that our proposed estimates are more efficient than Lee's in all the situations, in terms of accuracy, RMSE and shorter length of confidence interval. When $\phi$ is far below $1$ (i.e. for $\phi=0.60$), SEMWiG produce slightly better results for small population. This discrepancy increases when $N$ is larger. In all other situations, performance of $\hat{N}_{tb}^{I}$ is better than SEMWiG in all the 12 populations except P1 for both the values of $\phi$. Precisely, except P1 and the populations with $\phi=0.60$, we found that \textit{$\hat{N}_{tb}^{I}$ $<$ SEMWiG $<$ Lee}, in terms of RMSE and \textit{SEMWiG $<$ $\hat{N}_{tb}^{I}$ $<$ Lee} in terms of length of the associated interval estimates of $N$. Further, \textit{invariance} property of the estimates obtained through our proposed integrated likelihood is verified as the estimates for true $N=500$ are 2.5 times higher than that for true $N=200$. Lastly, another important feature of $\hat{N}_{tb}^{I}$ is that being a non-Bayesian pseudo-likelihood based inferential strategy, it does not incur serious computational effort as for existing Bayesian strategies - Lee's and SEMWiG.

\subsection{Real data Example I: Children Injury Data}
In Epidemiological study, use of capture-recapture experiment is very popular but more than two lists are hardly ever found. The simple estimate $\hat{N}_{ind}$ assuming list-independence is widely employed in this domain, even sometimes without judging its relevancy. Here we consider a work by Jarvis et al. (2000), in which authors illustrate the serious drawbacks in the use of this estimator specifically for injury related data. The problem was to get the count of children under 15 years of age from addresses in Northumbria who were seriously injured in local Motor Vehicles Accidents (MVA) between 1 April, 1990 and 31 March, 1995. One source was \textit{Stats19} data covering all road traffic accidents in Northumbria causing injuries to children that had been reported to the police and another was the \textit{Hospital Episode} data (HES) covering admissions of children. The associated DRS data are presented in Jarvis et al. (2000, Table 4, pp. 48) for three different classes - Cyclists, Passengers and Pedestrians. Jarvis et al. argued that children injured in MVAs as pedestrians or cyclists rarely enter insurance claims for which they have to inform police for case diary. Sometimes the police, in establishing whether an
injury is serious, are recommended to contact the hospital to find out whether the child is admitted or not. The common estimates under independence ($\hat{N}_{ind}$) for these three classes are shown in third column of Table \ref{tab:4}. It is noted that $\hat{N}_{ind}$'s are more than twice the total number of cases actually observed ($x_0$). Also, value of the estimate $\hat{c}$ for these three classes are 0.25, 0.40 and 0.59 respectively, which are substantially small. All these direct to the possibility of list dependency (indicating \textit{recapture aversion}, due to very small amount of recapture) and this motivate us to include this example in our illustration. These three classes have a common feature that $x_{1\cdot}<x_{\cdot1}$ (the next example of Handloom Data has the opposite feature). We present summary of results in Table \ref{tab:4} for our proposed integrated likelihood method along with Lee et al's Bayes and Chatterjee and Mukherjee's empirical Bayes SEMWiG estimates for comparison.
\begin{table}[h]
	\centering
	\caption{Summary results of the proposed integrated likelihood estimate along with independence estimate, Lee's full Bayes estimate and SEMWiG empirical Bayes estimates for \textbf{Children Injury Data}.}
	\begin{minipage}{14cm}
		\begin{tabular}{|lrcccc|}
			\hline
			Class $\&$ & &  & &  & \\
			(Estimate $\hat{c}$) & & $\hat{N}_{ind}$ & Lee & SEMWiG & $\hat{N}_{tb}^{I}$\\
			\hline
			\hline
			& & & & & \\
			Cyclists\footnote{Methods developed in Chatterjee and Mukherjee (2016c, pp. 5-6) suggest \textit{recapture aversion}} &  $\hat{N}$(s.e.) & 495 (69.68)&  292 (68.83) & 244 (19.99) & 303 (20.04)\\
			(0.254) &  C.I. & $(359, 632)$ & $(212, 448)$ & $(226, 273)$  &$(272, 350)$\\
			& & & & & \\
			Passengers$^{a}$  & $\hat{N}$(s.e.) & 249 (24.05) & 193 (21.85) & 181 (17.76) & 192 (9.05)\\
			(0.40) & C.I. & $(202, 296)$ & (163, 248) & $(170, 208)$  & $(175, 212)$\\
			& & & & & \\
			Pedestrians$^{a}$ & $\hat{N}$(s.e.) & 1323 (31.90) & 1213 (99.69) & 1110 (31.45)& 1159 (13.53)\\
			(0.592)  & C.I. & $(1260, 1385)$ & $(1050, 1424)$ & $(1071, 1148)$ &$(1135, 1186)$\\
			\hline
		\end{tabular}
	\end{minipage}
	\label{tab:4}
\end{table}

For the data on all the classes in Table \ref{tab:4}, all the competing estimators (Lee, SEMWiG \& $\hat{N}_{tb}^{I}$) agree with the negative departure from independence, i.e. recapture aversion and the resulting dual system estimates $\hat{N}_{ind}$ produce large estimates which seriously overestimates $N$. Lee's estimates posses larger variation than all other estimates and hence its credible intervals are too wide. Estimate $\hat{N}_{tb}^{I}$ has better efficiency than all other estimates. The proposed $\hat{N}_{tb}^{I}$ with the suggested directional knowledge (i.e. $\phi<1$) produces estimate between Lee and SEMWiG. Further, in most of the all cases, it has smaller variance and tighter confidence bounds, as expected.

\subsection{Real data Example II: Handloom Data}
Let us consider a new data from a survey aimed to estimate the undercount in the census of handloom workers residing at Gangarampur in South Dinajpur district of state West Bengal in India. The survey was post enumeration type (i.e. PES) and conducted in November 2013, three months after the census operation (\textit{see} SOSU (2014)). This data sets is also used in Chatterjee and Mukherjee (2016c) for real data illustration of their proposed behavioral dependence classification methods. Their classification strategies reveal that Ward 2 and Ward 16 has the nature of recapture proneness and aversion respectively. For details on the associated data sets in DRS format and possible threat of list-dependence through behavioral response variation, we refer to Chatterjee and Mukherjee (2016c). Being quite assured about the homogeneity within wards from the experts of Textile Directorate, we apply the model $M_{tb}$ for these data and compute the estimates following our proposed integrated likelihood method (in section \ref{Sec:Pro_Int_lik_Meth}). Here also we also compute Lee's Bayes estimate and the empirical Bayes estimate - SEMWiG for comparison in dependence situation. We also report the summary results if list-independence is assumed in order to measure the extent of deviation of other dependent estimates from independence.
\begin{table}[h]
	\caption{Summary results of the proposed integrated likelihood estimate along with independence estimate, Lee's full Bayes estimate and empirical Bayes estimate - SEMWiG for \textbf{Handloom Data}.}
	\begin{minipage}{40cm}
		\begin{tabular}{|lrcccc|}
			\hline
			Population $\&$ & &  & &  & \\
			(Estimate $\hat{c}$) & & $\hat{N}_{ind}$ & Lee & SEMWiG & $\hat{N}_{tb}^{I}$\\
			\hline
			\hline
			& & & &  & \\
			Ward 2\footnote{Chatterjee and Mukherjee (2016c, pp.10) suggests \textit{recapture proneness} for this population} & $\hat{N}$(s.e.) & 159 (4.50) & 187 (15.23) & 168 (1.52) & 164 (6.46)\\
			(0.675)  & C.I. &  $(150, 167)$ & $(160, 220)$ & $(166, 169)$ &$(154, 178)$\\
			& & & &  & \\
			Ward 16\footnote{Chatterjee and Mukherjee (2016c, pp.10) suggests \textit{recapture aversion} for this population} & $\hat{N}$(s.e.) & 270 (21.53) & 230 (37.56) & 211 (9.67) & 213 (7.95)\\
			(0.382) & C.I. &  $(228, 312)$ & $(185, 318)$ & $(202, 225)$ & $(200, 230)$\\
			\hline
		\end{tabular}
	\end{minipage}
	\label{tab:5}
\end{table}

Table \ref{tab:5} says that the proposed $\hat{N}_{tb}^{I}$ finds around 164 handloom workers assuming $\phi>1$. For the other sampled ward, when we incorporate the recapture aversion suggestion in our proposed method, it implies that approximately 213 workers are residing, which is very close to the SEMWiG estimate. Under the consideration of \textit{recapture proneness}, efficiency of proposed estimator is better than Lee's but little smaller than SEMWiG. When \textit{recapture aversion} is assumed, $\hat{N}_{tb}^{I}$ is most efficient.

\section{CONCLUSIONS}\label{conclu}
List-independence assumption does not hold satisfactorily in many instances. Various data from epidemiological studies and undercount or overcoount in demographic data motivate us to use a suitable model by avoiding the assumption of list-independence. As far as homogeneous human population size estimation is concerned, two-sample capture-recapture experiment is appropriate along with $M_{tb}$ modelling. Here, we consider integrated likelihood method as a non-Bayesian strategy which has a potential to overcome the non-identifiability in $M_{tb}$-DRS model. We have shown that general integrated likelihoods using common non-informative priors fail to produce estimate for population size $N$. To overcome this shortcoming, here we proposed an integrated likelihood based on suitable prior for \textit{unrelated} nuisance parameter with the help from a novel idea by Severini (2007). This pseudo-likelihood mechanism produces efficient estimates satisfying several properties including invariance, less prior sensitiveness, etc. To get rid of the aforesaid nonidentifiability problem, choice of suitable informative prior is always a statistical challenge. However in many instances (e.g. epidemiology or demography or economic), direction of the underlying behavioral dependency (i.e., whether the given population is likely to be \textit{recapture prone} or \textit{averse}) can be anticipated correctly. Therefore, choice of hyperparameters in priors are suggested depending upon the availability of such directional knowledge on $\phi$. Indeed, as per our knowledge, this article presents the first efficient non-Bayesian strategy for the complex $M_{tb}$-DRS model. Our simulation study supports the fact that this newly developed pseudo-likelihood method is either more efficient in some regular situations than the existing methods or it is comparable to them in other situations. Hence, this integrated likelihood method can be treated as an efficient and application worthy alternative estimation mechanism, which of course incurs very less computation burden than other existing methods for estimating $N$ in this context.


\section*{Appendix}
\begin{proof}{Proof of Theorem \ref{Theo_1}.}{}
The log-likelihood of the model $M_{tb}$ with parameterization in (\ref{Eq_L_1}) is\\
$\ell(\theta,\psi)=\ell(N,p_{1\cdot},c,p)=\Sigma_{i=0}^{x_0-1}ln(N-i)+x_{1\cdot}ln(p_{1\cdot})+(N-x_{1\cdot})ln(1-x_{1\cdot})+x_{01}ln(p)+(N-x_{0})ln(1-p)+x_{11}ln(c)+x_{10}ln(1-c)$ and, therefore
\begin{eqnarray}
\ell_{p_{1\cdot}}(N,p_{1\cdot},c,p)&=&(x_{1\cdot}/p_{1\cdot})-(N-x_{1\cdot})/(1-p_{1\cdot}),\nonumber\\
\ell_{c}(N,\psi_1,\psi_2)&=&\frac{x_{11}}{c}-\frac{x_{10}}{1-c}\nonumber\\ \ell_{p}(N,p_{1\cdot},c,p)&=&(x_{01}/p)-(N-x_{0})/(1-p).\nonumber
\end{eqnarray}
Now, if we take expectations on the partial derivatives of log-likelihood functions over the distribution fixing ($\theta,\psi$) at ($\theta_0,\psi_0$), we would have
\begin{eqnarray}
E(\ell_{p_{1\cdot}}(N,p_{1\cdot},c,p):N_0,p_{1\cdot;0},c_0,p_0)&=&\frac{N_0p_{1\cdot;0}}{p_{1\cdot}}-\frac{N-N_0p_{1\cdot;0}}{1-p_{1\cdot}},\label{Appen_01}\\
E(\ell_{c}(N,p_{1\cdot},c,p):N_0,p_{1\cdot;0},c_0,p_0)&=&\frac{N_0p_{11;0}}{c}-\frac{N_0(p_{1\cdot;0}-p_{11;0})}{1-c},\label{Appen_02}\\
E(\ell_{p}(N,p_{1\cdot},c,p):N_0,p_{1\cdot;0},c_0,p_0)&=&\frac{N_0p_{01;0}}{p}-\frac{N-N_0(p_{1\cdot;0}+p_{01;0})}{1-p}.\label{Appen_03}
\end{eqnarray}

Using Equation (\ref{Eq_IMLE_1}) we obtain the following result for strongly unrelated nuisance parameters ($\gamma_1,\gamma_2,\gamma_3$) from Equations (\ref{Appen_01})-(\ref{Appen_03}) respectively as\\
$\gamma_1=(N/\hat{N}_{ind})p_{1\cdot},\hspace{0.2in}\gamma_2=c,\hspace{0.2in} \gamma_3=\frac{p(1-p_{1\cdot})}{(\hat{N}_{ind}/N)-p_{1\cdot}}.$
\end{proof}

\begin{proof}{Proof of Result \ref{Res_1}.}{}
Nuisance parameters $\psi=(p_{1\cdot}, c, p)$ of the model $M_{tb}$ are expressed by the unrelated parameters $\gamma$, stated in Theorem \ref{Theo_1}, as follows:
\begin{eqnarray}
p_{1\cdot}&=&(\hat{N}_{ind}/N)\gamma_1,\label{Appen_04}\\
c&=&\gamma_2,\label{Appen_05}\\
p&=&\frac{\gamma_3(1-\gamma_1)}{(N/\hat{N})-\gamma_1}.\label{Appen_06}
\end{eqnarray}
Now, replacing $\psi=(p_{1\cdot}, c, p)$ in likelihood (\ref{Eq_L_1}) by $\gamma$ using equations (\ref{Appen_04})-(\ref{Appen_06}), we have rewritten the likelihood (\ref{Eq_L_1}) as
\begin{eqnarray}
L_{tb}(N,\gamma) &\propto& \frac{N!(1-\gamma_2)^{x_{10}}\gamma_2^{x_{11}}\gamma_1^{x_{1\cdot}}}{(N-x_0)!N^{x_{1\cdot}}}\left[\frac{\gamma_3(1-\gamma_1)}{(N/\hat{N})-\gamma_1} \right]^{x_{01}}\left(1-\frac{\hat{N}_{ind}}{N}\gamma_1\right)^{N-x_{1\cdot}}\times\nonumber\\
&&\left[1-\frac{\gamma_3(1-\gamma_1)}{(N/\hat{N})-\gamma_1}\right]^{N-x_{0}},\nonumber
\end{eqnarray}
provided $\gamma_1<(N/\hat{N})$. In order to proceed for integrated likelihood, we start by choosing Beta distributions for original nuisance parameters $p_{1\cdot}$ and $p$ with parameters ($r_1,s_1$) and ($r_2,s_2$) respectively, for positive real numbers $r_1$, $s_1$, $r_2$, $s_2$. Therefore, newly determined unrelated nuisance parameters $\gamma_1$ and $\gamma_3$ would follow the distributions $\pi(\gamma_1)\equiv GB1(b_1=\frac{N}{\hat{N}_{ind}},r_1,s_1)$ and $\pi(\gamma_3)\equiv GB1(b_2=\frac{1-p_{1\cdot}}{(\hat{N}_{ind}/N)-p_{1\cdot}},r_2,s_2)$ respectively, where GB1() stands for \textit{Generalized Beta distribution of Type 1}. We also consider the prior on $c=\gamma_2$ as $\pi(\gamma_2)\equiv Unif(0,1)$. So, $\pi(\gamma)$ is of the form $\pi(\gamma)=\pi(\gamma_1)\pi(\gamma_2)\pi(\gamma_3|\gamma_1)$.\\
Hence, following some algebraic simplification $L(\theta,\gamma|\textbf{\underline{x}})\pi(\gamma)$ in Equation (\ref{Eq_05}) becomes
\begin{eqnarray}
L_{tb}(N,\gamma|\textbf{\underline{x}})\pi(\gamma) &\propto& \frac{N!(N-x_0+s_2-1)!}{(N-x_0)!(N-x_{1\cdot}+r_2+s_2-1)!}  \frac{\gamma_1^{x_{1\cdot}+r_1-1}}{N^{x_{1\cdot}+r_1}}\times\nonumber\\
&&\left(1-\frac{\hat{N}_{ind}}{N}\gamma_1\right)^{N-x_{1\cdot}+s_1-1}\label{Appen_08}
\end{eqnarray}
with $\theta=N$. Now, if we integrate Equation (\ref{Appen_08}) w.r.t. $\gamma$, then we have
\begin{eqnarray}
\overline{L}_{tb}^{I}(N) &\propto& \frac{N!(N-x_0+s_2-1)!(N-x_{1\cdot}+s_1-1)!}{(N-x_0)!(N-x_{1\cdot}+r_2+s_2-1)!(N+r_1+s_1-1)!}\nonumber\\ &=&\frac{\Gamma(N-x_0+s_2)\Gamma(N+1)}{\Gamma(N+r_1+s_1)\Gamma(N-x_0+1)},\nonumber
\end{eqnarray}
assuming $r_2+s_2=s_1$. Thus, we have Equation (\ref{Eq_IMLE_3}).\\ 
Now, 
\begin{eqnarray}
\frac{\overline{L}_{tb}^{I}(N+1)}{\overline{L}_{tb}^{I}(N)} &\propto& \frac{(N-x_0+s_2)...(N-x_0+2)}{(N+r_1+s_1)...(N+2)}
\times \frac{(N+r_1+s_1-1)...(N+1)}{(N-x_0+s_2-1)...(N-x_0+1)}\nonumber\\
&=& \frac{(N-x_0+s_2)(N+1)}{(N+r_1+s_1)(N-x_0+1)}.\nonumber
\end{eqnarray}
Therefore, 
\begin{eqnarray}
\frac{\overline{L}_{tb}^{I}(N+1)}{\overline{L}_{tb}^{I}(N)} &\geq& 1\nonumber\\ 
\Leftrightarrow s_2+s_2N-x_0 &\geq& (r_1+s_1)N + (r_1+s_1)(1-x_0)\nonumber\\
\Leftrightarrow (N+1)(r_1+s_1-s_2)  &\leq& x_0(r_1+s_1-1)\nonumber\\
\Leftrightarrow N  &\leq& \frac{x_0(r_1+s_1-1)}{(r_1+s_1-s_2)}-1.\nonumber
\end{eqnarray}
Hence the result in Equation (\ref{Eq_09}) since $r_2+s_2=s_1$. $\Box$
\end{proof}

\begin{proof}{Proof of Theorem \ref{Theo_2}.}{}
For $r_1=x_{1\cdot}s_1/(N-x_{1\cdot})=r_1(N, s_1)$, $\exists$ a real number, say $N_0=N_0(\underline{\textbf{x}},s_1,r_2)$, for which inequality in Equation (\ref{Eq_09}) is equivalent to $N\leq N_{0}<\infty$.
\begin{eqnarray}
N  &\leq& \frac{x_0(r_1+s_1-1)}{(r_1+r_2)}-1\nonumber\\
\frac{N+1}{x_0} &\leq& \frac{Ns_1-N+x_{1\cdot}}{Nr_2+x_{1\cdot}(s_1-r_2)}\nonumber
\end{eqnarray}
Therefore, the above inequality simplifies to
\begin{eqnarray} N^2r_2-N(x_{1\cdot}r_2+x_{01}s_1-r_2-x_{0})-x_{1\cdot}(x_{0}+r_2-s_1)&\leq& 0.\label{Eqn01_Th02}
\end{eqnarray}
Hence there exists two roots of the quadratic equation (\ref{Eqn01_Th02}), say $N'_0$ and $N_0$, both functions of ($\underline{\textbf{x}},s_1,r_2$) such that $N'_0\leq N \leq N_0.$\\ 
Now, the constant part of the quadratic function in the left side of the inequality (\ref{Eqn01_Th02}) remains positive iff $b<\frac{x_0}{N^*-x_0}$, since $r_2-s_1=-s_2=-b(N^*-x_0)$ (see section \ref{Imple_Lik01}). Hence, the smaller root $N'_0$ becomes negative iff $b<\frac{x_0}{N^*-x_0}$. Unless $\frac{x_0}{N^*-x_0}\leq1$ or equivalently, $N^*\geq2x_0$, $N'_0$ will be negative for both of the choices of $b$ values mentioned in section \ref{Imple_Lik01}. $\Box$
\end{proof}


\begin{thebibliography}{40}
\bibitem{Bish75} Bishop, Y., Fienberg, S. and Holland, P.(1975), \textit{Discrete Multivariate Analysis, Theory and Practice}, Cambridge, Massachusetts: MIT Press.
\bibitem{Bolfarine92} Bolfarine, H., Leite, J. G. and Rodrigues, J.(1992), \textit{On the Estimation of the Size of a Finite and Closed Population}, Biometrical Journal \textbf{34}, 577-593.
\bibitem{Chandra49} ChandraSekar, C. and Deming, W. E. (1949). On a method of estimating birth and death rates and the extent of registration. JASA \textbf{44}, 101-115.
\bibitem{Chao00} Chao, A., Chu, W. and Chiu, H.H.(2000), \textit{Capture-Recapture when Time and Behavioral Response Affect Capture Probabilities}, Biometrics \textbf{56}, 427-433.
\bibitem{Chatterjee16a} Chatterjee, K. and Mukherjee, D. (2016a). An Improved integrated likelihood estimator population size estimation in dual record system. \textit{Statistics and Probability Letters} \textbf{110}, 146-154.
\bibitem{Chatterjee16b} Chatterjee, K. and Mukherjee, D. (2016b). On the Estimation of Homogeneous Population Size from a Complex Dual-record System. \textit{Journal of Statistical Computation and Simulation} \textbf{86}, 3562-3581.
\bibitem{Chatterjee16c} Chatterjee, K. and Mukherjee, D. (2016c). On the identification of the nature of behavioural dependence in a two-sample capture-recapture study. Technical Report. ResearchGate DOI: 10.13140/RG.2.1.2146.7926
\bibitem{Cox87} Cox, D. R. and Reid, N. (1987). \textit{Parameter orthogonality and approximate conditional infer-
	ence (with discussions)}, J. R. Statist. Soc. B \textbf{49}, 1-39.
\bibitem{George92} George, E. I. and Robert, C. P.(1992), \textit{Capture-recapture estimation via Gibbs sampling}, Biometrika, \textbf{79}, 677-683.
\bibitem{Gosky11} Gosky, R. and Ghosh, S. K. (2011). A Comparative Study of Bayes Estimators of Closed Population Size from Capture-Recapture Data. \textit{Journal of Statistical Theory and Practice} \textbf{5}, 241-260.
\bibitem{Hugg89} Huggins, R.(1989), \textit{On the statistical analysis of capture-recapture experiments}, Biometrika, \textbf{76}, 133-140.
\bibitem{Jarvis00} Jarvis, S. N., Lowe, P. J.; Avery, A., Levene, S., Cormack, R. M. (2000), \textit{Children are not goldfish ? mark/recapture techniques and their application to injury data}, Injury Prevention \textbf{6}, 46-50.
\bibitem{Lee03} Lee, S. M., Hwang, W.H. and Huang, L.H.(2003), \textit{Bayes estimation of Population Size from Capture-recapture Models with Time Variation and Behavior response}, Statistica Sinica, \textbf{13}, 477-494.
\bibitem{Lloyd94} Lloyd, C.J.(1994), \textit{Efficiency of martingale methods in recapture studies}, Biometrika, \textbf{81}, 305-315.
\bibitem{Nour82} Nour, E. S. (1982), \textit{On the Estimation of the Total Number of Vital Events with Data from Dual-record Collection Systems}, J. R. Statist. Soc. A \textbf{145}, 106-116.
\bibitem{Otis78} Otis, D.L., Burnham, K.P., White, G.C. and Anderson, D.R.(1978), \textit{Statistical Inference from Capture Data on Closed Animal Populations}, Wildlife Monographs, \textbf{62}, 1-135.
\bibitem{Salasar14} Salasar, L. E. B., Leite, J. G. and Louzada, F. (2015). On the integrated maximum likelihood estimators for a closed population capture?recapture model with unequal capture probabilities. \textit{Statistics} \textbf{49}, 1204-1220. 
\bibitem{Severini00} Severini, T.A. (2000). \textit{Likelihood Methods in Statistics}, Oxford University Press Inc., New York.
\bibitem{Severini07} Severini, T.A. (2007). Integrated Likelihood Functions for Non-Bayesian Inference. \textit{Biometrika} \textbf{94}, 529-542.
\bibitem{SOSU14} SOSU (2014). Report on the Project "\textit{Survey of Looms and Work sheds in Comprehensive Handloom Development Programme in Dakshin Dinajpur district}? by Sampling and Official Statistics Unit, Indian Statistical Institute, Commissioned by: Directorate of Textiles, Government of West Bengal, India, 5th March 2014.
\bibitem{Wang15} Wang, X., He, C. Z. and Sun, D. (2015). Bayesian Estimation of Population Size via Capture-Recapture Model with Time Variation and Behavioral Response. \textit{Journal of Ecology} \textbf{5}, 1-13.
\bibitem{Wolter86} Wolter, K. M. (1986). Some Coverage Error Models for Census Data. \textit{JASA} \textbf{81}, 338-346.
\bibitem{Xu14} Xu, C., Sun, D. and He, C. (2014). Objective Bayesian analysis for a capture-recapture model. \textit{Ann Inst Stat Math} \textbf{66}, 245?278.
\bibitem{Yang05} Yang, H. C. and Chao, A. (2005). Modelling Animals' Behavioral Response by Markov Chain Models for Capture-Recapture Experiments. \textit{Biometrics} \textbf{61}, 1010-1017.
\end{thebibliography}
\end{document}